\documentclass[preprint,aps,showpacs,onecolumn]{revtex4}
\usepackage{amsfonts}
\usepackage{amsmath}
\usepackage{amssymb}
\usepackage{graphicx}
\usepackage[latin1]{inputenc}

\newcommand{\be}{\begin{equation}}
\newcommand{\ee}{\end{equation}}
\newcommand{\bes}{\begin{subequations}}
\newcommand{\ees}{\end{subequations}}
\newcommand{\ben}{\begin{eqnarray}}
\newcommand{\een}{\end{eqnarray}}

\begin{document}

\title{Kink-antikink collisions for twin models}
\author{A. R. Gomes$^{1}$, R. Menezes$^{2,3}$, K. Z. Nobrega ${^4}$, F. C.
Simas$^{5}$}
\affiliation{$^1$ Departamento de F\' isica, Instituto Federal do Maranh\~ao (IFMA),
Campus Monte Castelo, 65030-005, S\~ao Lu\'\i s, Maranh\~ao, Brasil\\
$^2$ Departamento de Ci\^encias Exatas, Universidade Federal da Para\'\i ba,
58297-000, Rio Tinto, Para\'\i ba, Brasil\\
$^3$ Departamento de F\' isica, Universidade Federal de Campina Grande,
58109-970, Campina Grande, Para\'\i ba, Brasil\\
$^4$ Departamento de Eletro-Eletr\^onica, Instituto Federal do Maranh\~ao
(IFMA), Campus Monte Castelo, 65030-005, S\~ao Lu\'\i s, Maranh\~ao, Brasil\\
$^5$ Departamento de F\'\i sica, Universidade Federal do Maranh\~ao (UFMA) \\
Campus Universit\'ario do Bacanga, 65085-580, S\~ao Lu\'\i s, Maranh\~ao,
Brasil }

\begin{abstract}
In this work we consider kink-antikink collisions for some classes of $(1,1)$%
-dimensional nonlinear models. We are particularly interested to investigate
in which aspect the presence of a general kinetic content in the Lagrangian
could be revealed in a collision process. We consider a particular class of
models known as twin theories, where different models lead to same solutions
for the equations of motion and same energy density profile. The theories
can be distinguished in the level of linear stability of defect structure.
We study a class of k-defect theories depending on a parameter $M$ which is
the twin theory of the usual $\phi^4$ theory with standard dynamics. For $%
M\to\infty$ both models are characterized by the same potential. In the
regime $1/M^2<<1$, we obtain analytically the spectrum of excitations around
the kink solution. It is shown that with the increasing on the parameter $%
1/M^2$: i) the gap between the zero-mode and the first-excited mode
increases and ii) the tendency of one-bounce collision between kink-antikink
increases. We numerically investigate kink-antikink scattering, looking for
the influence of the parameter changing for the thickness and number of
two-bounce windows, and confronting the results with our analytical findings.
\end{abstract}

\maketitle


\section{Introduction}

In this work we consider kink-antikink collisions for some classes of $(1,1)$%
-dimensional nonintegrable models. As one knows, the collision process for
integrable systems has an intrinsic simplicity, with the solitary waves
passing to each other with at most a phase shift. Despite this simplicity,
analytical results for most integrable models is rather nontrivial. As an
example, the analysis of kink-kink and kink-antikink scattering for
integrable models using perturbative analysis can be found in Refs. \cite%
{scat1,scat2}. For more recent results on this subject, see Refs. \cite%
{scat3,scat4,scat5}. The effect of a phase shift of the collision process
for an integrable model can be confronted to the richness of the collision
process for nonintegrable models, where most of analysis must be done
numerically. As already shown by Anninos et al \cite{aom} for the $%
\lambda\phi^4$ theory, for sufficiently small initial velocities, the kink
and antikink capture one another and a trapped bion state is formed. On the
other hand for larger velocities a simple collision occurs, and after the
contact, the pair of defects retreats from each other. For intermediate
initial velocities, the structure of the collision is fair more complex,
with two-bounce windows appearing in an endless sequence between the larger
bion region and the one-bounce ones. A self-similar structure is revealed if
one zooms the initial approximation velocity close to the interface regions.
Indeed, if one refines the initial velocity near to an interface between a
two-bounce window and a bion window, one sees the appearance an endless
sequence of three-bounce and bion windows. This refinement process continues
with higher number of bounces being observed; however, a limit in this
process is achieved due to losses by radiation. The higher number of bounces
appears jointly with a higher number of internal mode oscillations. The
different windows are related by a scaling relation between the window
velocity thickness and the number of internal mode oscillations.

The intriguing character of these alternating regions conflicts with the
naive expectation that bion states could not be formed for larger initial
velocities than that verified for n-bounce collisions. A
semi-phenomenological theory that accounts for the presence of two-bounce
windows is due to Campbell et al \cite{csw}. In their work, the two-bounce
behavior is described as a two-steps interacting process. In the first
interaction the energy is transferred from the translational mode to an
internal shape mode oscillation of the kink. In the second one, if the kink
and antikink satisfy a resonance criteria \cite{csw,pc}, the energy in the
vibrational mode is turned to the translational mode, and the pair $K\bar K$
is liberated from their mutual attraction. Quantitatively the simple
relation must be obeyed \cite{csw} 
\begin{equation}  \label{T_omega}
\omega_1 T=2n\pi+\delta
\end{equation}
where $\omega_1$ is the frequency of the internal mode, T is the time
interval between the bounces, $n$ is an integer and $\delta$ is the phase
shift between the incoming and outgoing kink. In this way, the higher is $%
\omega_1$, the smaller is the time interval between the bounces, which
signals that the energy transference from translational to vibrational mode
is more difficult to be achieved. The simple one bounce scattering process
occurs for the initial kink velocity higher than a critical velocity $v_c$.
A heuristic expression for the relation between $T$ and $v$ for the
two-bounce windows is presented also in \cite{bk}, namely 
\begin{equation}
T\propto \sqrt{v_c-v}, \,\,\,v<v_c.
\end{equation}
The former two expressions can be used to understand quantitatively how the
windows centers scale with $n$ \cite{bk}, with the successive windows with
even smaller thickness accumulating near $v=v_c$.

The escape of the pair kink-antikink from their mutual attraction is
verified when the energy of the vibrational mode is less than the kinetic
energy of the colliding kinks \cite{bk}. Usually one can use collective
coordinates to obtain the $K\bar K$ attractive interaction potential $%
U_{K\bar K}$ as a function of the separation of the pair kink-antikink \cite%
{kudr,sug}. The potential $U_{K\bar K}$ can be intuitively understood as the
energy of the static field configuration consisting of a kink at $+Z$ and an
antikink at $-Z$ \cite{cps}. On the other hand, if the time duration of
internal oscillations is accompanied by a leaking of energy by radiation
greater than the kinetic energy of the colliding kinks, a trapped bion $%
K\bar K$ state is formed \cite{bk}.

The analysis of kink-antikink collisions can be found in the literature for
several interesting examples of nonintegrable models. In addition to the
already cited $\lambda\phi^4$ model \cite{aom}, one can cite the modified
sine Gordon \cite{pc} and the $\phi^6$ model \cite{dmrs}. Most of
nonintegrable models have internal oscillatory modes, responsible for
resonant scattering. However, there are exceptions, as the $\phi^6$ model
where despite the absence of an internal oscillatory mode, resonant
scattering was reported \cite{dmrs}. There, the potential for linear
perturbations has a wide central well and the energy can be transferred from
the translational mode to an extended meson state residing in this potential 
\cite{dmrs}. Indeed, the central well allow several discrete eigenvalues
corresponding to meson-soliton bound states \cite{lohe}. This is contrary to
what happens for the more usual models $\phi^4$ and sine-Gordon where mesons
can pass through the kink and antikink without reflection, with only a phase
shift \cite{lohe}. In the $\phi^6$ model the two-bounce windows satisfy the
same relation given by eq. (\ref{T_omega}), but with $\omega_1$ as the
frequency of the lowest collective mode. Some interesting discussions
related to this topic can also be seen in Ref. \cite{weigel}.

Our proposal here is to investigate in which measure the kink-antikink
collision processes can be used to distinguish twin models, i.e. a class
of topological defects with the same scalar field profiles and energy
densities \cite{altw}. For some recent results on this subject, see \cite%
{br, bhm,bllm}. We are particularly interested in k-defect theories, in part due
to their use for explaining the accelerated expansion of the universe \cite%
{k_cosm}. Kink-antikink collision processes are useful for cosmology in
theories with one extra dimension \cite{tm}, as in the ekpyrotic proposal
for brane collision \cite{ekpy}. Also, one can cite theories in the
braneworld scenario with generalized dynamics \cite{bdglm, blom}. The study
of bubble cosmology has now being a subject of renewed interest since it can
lead to possible ways to probe string landscape \cite{pol}. Cosmic bubble
collisions in the regime of high nucleation rate can be studied ignoring the
expansion and the curvature of the universe \cite{haw}. Despite the usual
approach being lattice simulations in $(3,1)$ dimensions \cite{egl}, in the
special situation of bubbles with $SO(2,1)$ symmetry, a high speed collision
of two bubbles is equivalent, depending on the vacuum configuration, to a $%
(1,1)$ dimensional $K\bar K$ collision \cite{lim}. This shows that the studies of $K\bar
K$ collision in twin theories can be useful for several cosmological
scenarios.

In this work, as a starting point we will focus on $(1,1)$-dimensional
scalar field theories in the Minkowski spacetime. Despite the simplicity of
the proposal, we will show that our results lead to interesting insights
about the influence of the k-dynamics on some characteristics of the
collision, namely the presence of two-bounce windows. For this purpose, in
the Sect. II we review the main first-order formalism for twin theories.
Stability analysis is reviewed in Sect. III. Sect. IV specializes the
discussion for the $\phi^4$ model and its twin theory. The discrete spectra
of fluctuations is obtained for mass parameter $1/M^2\ll1$ and compared with
known results for the $\phi^4$ theory. It is shown that an increasing of the
parameter $1/M^2$ increases the gap between the zero- and excited modes.
This is confronted with our numerical results, presented in Sect. V. Our
main conclusions are presented in Sect. VI.


\section{Twin theories}


We start revising some results of the general formalism of k-defects in $%
(1,1)$ dimensions \cite{bab,asw,blmo}. 
Consider a general Lagrangian density
given by 
\begin{equation}
\mathcal{L}=\mathcal{L}(\phi,X)
\end{equation}
with 
\begin{equation}
X=\frac12 \partial_\mu \phi \partial^\mu \phi.
\end{equation}
A remark about notation: in this paper a subscript in $\mathcal{L}$ (and later on, in $W$) means partial derivative with respect to the argument. Then for instance $\mathcal{L}_X\equiv\partial\mathcal{L}/\partial X$, $\mathcal{L}_{X\phi}\equiv\partial^2\mathcal{L}/(\partial X\partial \phi)$ and so on. Now, the equation of motion is given by 
\begin{equation}
\partial_\mu \left(\mathcal{L}_X \partial^\mu \phi\right)=\mathcal{L}_\phi
\end{equation}
or 
\begin{equation}  \label{eqm1}
G^{\mu\nu}\partial_\mu\partial_\nu \phi = - 2X \mathcal{L}_{X\phi}+ \mathcal{%
L}_{\phi}
\end{equation}
with 
\begin{equation}
G^{\mu\nu}=\mathcal{L}_X \eta^{\mu\nu} +\mathcal{L}_{XX} \partial^\mu \phi
\partial^\nu \phi
\end{equation}
The energy-momentum tensor is 
\begin{equation}
T^{\mu\nu}=\mathcal{L}_X \partial^\mu \phi \partial^\nu \phi- \eta^{\mu \nu}%
\mathcal{L}
\end{equation}
and the corresponding energy is 
\begin{equation}
E=\int^{\infty}_{-\infty} T^{00} \, dx.
\end{equation}
For static solutions, $\phi=\phi(x)$, and Eqs. \eqref{eqm1} can be rewritten
as 
\begin{equation}  \label{eqmest}
\mathcal{L}_X A^2\phi^{\prime\prime} =- 2X\mathcal{L}_{X\phi}+ \mathcal{L}%
_\phi
\end{equation}
with 
\begin{equation}  \label{Asq}
A^2=\frac{2 X \mathcal{L}_{XX}+\mathcal{L}_X}{\mathcal{L}_X}.
\end{equation}

Here we will be interested in the corresponding model in the modified
k-defect theory of a scalar field model with the standard Lagrangian density 
\begin{equation}  \label{L_st}
\mathcal{L}_{(S)}=X-V(\phi)
\end{equation}
where $V(\phi)$ is the potential for the standard theory. The equation of
motion for static solutions gives 
\begin{equation}
\phi^{\prime \prime }=V_{\phi}
\end{equation}
and the energy density is 
\begin{equation}
T^{00}\equiv\rho_S(x)=\frac12{\phi^{\prime }}^2+V(\phi).
\end{equation}
A very interesting proposal for a corresponding k-defect theory can be found
in \cite{altw}: 
\begin{equation}  \label{L_kt}
\mathcal{L}_{(m)}=M^2-M^2\biggl(1+\frac{U(\phi)}{M^2} \biggr)\sqrt{1-\frac{2X}{%
M^2}}.
\end{equation}
where $M$ is a mass scale. As pointed out in \cite{bdglm}, the limit $%
1/M^2\to0$ turns this Lagrangian in a standard form with potential $U(\phi)$%
. In this way we can say $U(\phi)$ is the potential of the modified model.
Stable static kinklike solutions must be pressureless ($T^{11}=0$) \cite%
{bdglm,blmo}. In this way the equation of motion gives 
\begin{equation}  \label{phi'U}
{\phi^{\prime }}^2=2U(\phi)+\frac{U^2(\phi)}{M^2}
\end{equation}
and the energy density is 
\begin{equation}
\rho_m(x)={\phi^{\prime }}^2.
\end{equation}
If the potentials for the standard and k-defect theory are related by \cite%
{altw} 
\begin{equation}  \label{VU_relation}
V(\phi)=U(\phi)+\frac12\frac{U(\phi)^2}{M^2}
\end{equation}
then the models of Eqs. (\ref{L_st}) and (\ref{L_kt}) have the same stable
defect structure $\phi(x)$ and energy density \cite{altw,bdglm}.

The first-order formalism for general Lagrangian density $\mathcal{L}%
(X,\phi) $ \cite{blm} is very useful to understand the connections between
the potentials. Introducing a function $W(\phi)$ for the standard Lagrangian
density, and for the potential 
\begin{equation}
V(\phi)=\frac12W_{\phi}^2
\end{equation}
the solutions of the second-order equation of motion are also solutions of 
\begin{equation}  \label{eom_st}
\phi^{\prime }=W_{\phi}.
\end{equation}
When applied to the k-defect theory, the corresponding potential is \cite%
{bdglm} 
\begin{equation}  \label{U_M}
U(\phi)=-M^2+M^2\sqrt{1+\frac{W_{\phi}^2}{M^2}}
\end{equation}


\section{Stability analysis}


Here we will review the linear stability of the solutions. We consider $%
\phi=\bar\phi+\eta$, where $\bar\phi$ is the unperturbed solution and we
suppose small perturbations $\eta$ around this solution. We can use Eq. %
\eqref{eqm1} to attain the first-order contribution in $\eta$ 
\begin{equation}  \label{eigenEq0}
\partial_\mu \!\!\left(\mathcal{L}_X \partial^\mu \eta +\mathcal{L}_{XX}
\partial^{\mu}\phi \partial_\alpha \phi \partial^\alpha\eta \right)\!=\! %
\left[\mathcal{L}_{\phi\phi}\!-\!\partial_\mu \!\left(\mathcal{L}_{\phi X}
\partial^\mu \phi\right)\right] \eta
\end{equation}

We decompose the perturbations in terms of the modes  
\begin{equation}
\eta (x,t)=\sum_{n}a_{n}\cos (\omega _{n}\,t)\eta _{n}(x)  \label{expan}
\end{equation}%
where the $a_{n}$ are real coefficients. Eq. (\ref%
{expan}) allow us to rewrite Eq. (\ref{eigenEq0}) as \cite{blmo} 
\begin{equation}
-\left[ A^{2}\mathcal{L}_{X}\eta _{n}^{\prime }\right] ^{\prime }=[\mathcal{L%
}_{\phi \phi }+(\mathcal{L}_{\phi X}\phi ^{\prime })^{\prime }+\omega
_{n}^{2}\mathcal{L}_{X}]\eta _{n},  \label{eigenEq}
\end{equation}%
and $A$ was defined in Eq. (\ref{Asq}). This is a Sturm-Liouville equation,
which means that the eigenfunctions $\eta _{n}$ satisfy a condition of
orthonormality with weight function $\mathcal{L}_{X}$ \cite{blmo}: 
\begin{equation}
\int_{-\infty }^{\infty }dx\,\mathcal{L}_{X}\,\eta _{n}(x)\eta
_{m}(x)dx=\delta _{mn},  \label{etaeta}
\end{equation}%
where the eigenfunctions $\eta _{n}$ obey appropriated
boundary conditions or $\mathcal{L}_{X}\,$ converges to zero more rapidly
than $\,\eta _{n}\left( x\right) $ at boundaries. The finiteness
of Eq. (\ref{etaeta}) must be analyzed in detail for every model to be
studied. 

{The eigenvalue problem (\ref{eigenEq}) becomes more clear after a
convenient change of variables (a procedure described in Ref. \cite{blm}) }%
\begin{equation}
dx=Adz,~~\eta _{n}=\frac{u_{n}}{\sqrt{\mathcal{L}_{X}A}},
\end{equation}%
{a Schrödinger-like equation can be obtained }%
\begin{equation}
-\left( u_{n}\right) _{zz}+U_{sch}(z)u_{n}=\omega _{n}^{2}u_{n},
\label{sch_eq}
\end{equation}%
{with \cite{blmo} }%
\begin{equation}
U_{sch}(z)=\frac{({A\mathcal{L}_{X}})_{zz}^{\frac{1}{2}}}{({A\mathcal{L}_{X}}%
)^{\frac{1}{2}}}-\frac{1}{{\mathcal{L}}_{X}}\biggl[\mathcal{L}_{\phi \phi }+%
\frac{1}{A}\biggl(\mathcal{L}_{\phi X}\frac{\phi _{z}}{A}\biggr)_{z}\biggr].
\end{equation}%
{Note that Eq. (\ref{sch_eq}) is an eigenvalue equation where stable
solutions correspond to }$\omega _{n}^{2}\geq 0${. The existence or
not of tachyonic modes (}$\omega _{n}^{2}<0${) depends on the model
and must be analyzed separately for each case. The determination of the
eigenvalues }$\omega _{n}$ {\ and stability analysis for the
particular case of the }$\phi ^{4}$ model and its twin counterpart
is considered in the next section.


\section{The $\protect\phi^4$ model and its twin counterpart}


In this section we will consider comparatively some properties of two
specific twin models. The set of bound states obtained from the stability
analysis will give informations that will be confronted later with the
numerical analysis of the collisions. It is important to note that, for the
models considered in this work, the absence of tachyonic modes is
demonstrated in Sect. II of Ref. \cite{bdglm}. Also, numerical analysis
discussing their eigenmodes can be found in two different ways in Refs. \cite%
{altw,bdglm}.

We start considering a standard Lagrangian density with a $\phi^4$
potential, where 
\begin{equation}
W_{\phi}=1-\phi^2.
\end{equation}
The kink-like solution of the first-order equation of motion (Eq. (\ref%
{eom_st})) is 
\begin{equation}
\phi_S(x)=\tanh(x-x_0)
\end{equation}
where $x_0$ is a constant, identified as the center of the kink. Note that,
as a consequence of the twin model construction, the former expression also
corresponds to the solution $\phi_m(x)$ achieved for the modified k-defect
theory. The first step in the investigation the kink-antikink collision
process is to investigate the spectra of fluctuations for both models. The
fluctuation modes around the kink are described as $\phi(x, t) = \phi_S(x) +
\sum_n a_n \eta_n(x) \cos(\omega_n t)$. For the standard Lagrangian, a
Schr\"odinger-like equation is attained 
\begin{equation}
-\eta_n^{\prime \prime }(x)+V_q(x)\eta_n(x)=\omega_n^2\eta(x)
\end{equation}
with the potential
\begin{equation}
V_q(x)=W_{\phi\phi}^2+W_{\phi}W_{\phi\phi\phi}=2\big(3\tanh^2(x)-1\big).
\end{equation}
For the twin k-defect model, a Schr\"odinger-like equation 
\begin{equation}
-u_{zz}+U_q(z)u=\omega^2u,
\end{equation}
is possible after a change of variables \cite{blm}: 
\begin{equation}
dx=\biggl(1+\frac{W_{\phi}^2}{M^2}\biggr)^{-\frac12} dz,
\end{equation}
\begin{equation}
\eta=\biggl(1+\frac{W_{\phi}^2}{M^2}\biggr)^{\frac14} u.
\end{equation}
Analytic expressions for the potential $U_q(z)$ can be attained in the
regime $1/M^2\ll1$. For the particular case $W_\phi=1-\phi^2$ considered,
one gets \cite{bdglm} 
\begin{equation}  \label{Uqz}
U_q(z) = 4 - 6 \mathrm{sech}^2(z) + \frac1{M^2} [4 \mathrm{sech}^2(z)+14 
\mathrm{sech}^4(z)-21 \mathrm{sech}^6(z)].
\end{equation}
We see that for $1/M^2\to0$ we have identical potentials $V_{(q)}(z)=U_{(q)}(z)$ for
the standard and k-defect theories, with the Schr\"odinger-like potential
being a modified P\"osh-Teller. This potential, analyzed in
Sugyama \cite{sug} in the context of $K\bar K$ collisions, has two discrete eigenvalues $\omega_i^2$ with corresponding
eingenfunctions $u_i(z)$: 
\begin{equation}
\omega_0^2=0, \,\,\, u_0(z)=\sqrt{\frac34}\mathrm{sech}(z)^2
\end{equation}
and 
\begin{equation}
\omega_1^2=3, \,\,\,u_1(z)=\sqrt{\frac32}\tanh(z)\mathrm{sech}(z),
\end{equation}
followed by a continuum of modes. The first excited state is an excitation
trapped in the kink, crucial for the formation of two-bounce states \cite%
{sug}. 
\begin{figure}[tbp]
\includegraphics[{angle=0,width=8cm,height=7cm}]{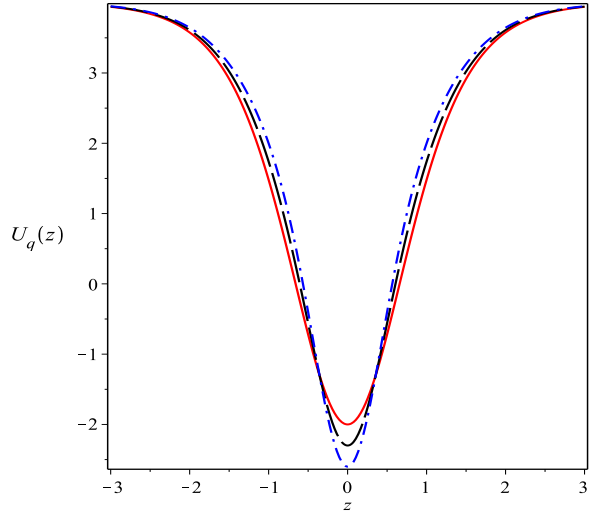} 
\caption{Schrodinger-like potential $U_q(z)$ for $1/M^2\to0$ (red line), $%
0.1 $ (black traced line) and $0.2$ (blue point-traced line).}
\label{fig_sch}
\end{figure}
Now, from Fig. \ref{fig_sch} we see that the increasing of $1/M^2$ leads to
the potentials $V_q(z)$ and $U_q(z)$ to depart one from another, with the
Schr\"odinger-like potential in the k-defect theory with a deeper and
thinner potential. Nondegenerate perturbation theory at first-order in the
parameter $1/M^2\ll 1$ can be used to attain the energy eigenvalues and
eigenfunctions of the twin k-defect theory. Eq. (\ref{Uqz}) can be written
as $U_q(z)=V_q(z)+U_{pert}(z)/M^2$, where the perturbation potential is 
\begin{equation}
U_{pert}=4 \mathrm{sech}^2(z)+14 \mathrm{sech}^4(z)-21 \mathrm{sech}^6(z).
\end{equation}
The ground state energy is invariant under such perturbation: 
\begin{eqnarray}
(\omega^{(1)}_0)^2 &=&\omega_0^2+\frac1{M^2}\int_{-\infty}^{\infty} dz |u_0(z)|^2 U_{pert}(z) = 0.
\label{omega0_pert}
\end{eqnarray}
This means the presence of the translation mode for the corresponding twin
model. The energy of the first-excited mode is changed to 
\begin{eqnarray}
(\omega^{(1)}_1)^2 &=&\omega_1^2+\frac1{M^2}\int_{-\infty}^{\infty} dz |u_1(z)|^2 U_{pert}(z)=3
+\frac85\frac1{M^2} \label{omega1_pert}.
\end{eqnarray}

\begin{table}[tbp]
\begin{tabular}{|c|c|c|c|}
\hline\hline
$1/M^2$ & $(\omega^{(1)}_1)^2$ & $(\omega^{(1)}_1)_{num}^2$ & relative error \\ 
\hline\hline
0 & 3 & 3.0000013 & $-4.3\times 10^{-7}$\\ \hline
0.01 & 3.016 &  3.0159773 & $7.5\times 10^{-6}$\\ \hline
0.1 & 3.16 & 3.1574820 & $8.0 \times 10^{-4}$ \\ \hline
\end{tabular}%
\caption{Energy of the first-excited state. In the second column we compare $(\omega^{(1)}_1)^2$, given by Eq. (\ref{omega1_pert}) obtained with time-independent perturbed theory. The third column shows the energy obtained numerically with a finite-element method. The fourth column shows the relative error between the results from second and third columns.}
\end{table}

Table I shows, for some values of $1/M^2$, the evaluated eigenvalue of the internal mode from Eq. (\ref{omega1_pert}) and the corresponding numerical solution of the Schr\"oedinger-like equation. We see from the Table that the relative error increases with $1/M^2$, with an acceptable value $\sim 10^{-5}$ for $1/M^2=0.01$. An increasing of $1/M^2$ by one order of magnitude increases the relative error $100$ times. Then, $1/M^2=0.1$ seems to be a too large value for the approximation to be valid.   Also, from Eqs. (\ref{omega0_pert}) and (\ref{omega1_pert}), we see that there is a larger gap between the zero-mode and the first
excited state with the increasing of the parameter $1/M^2$. This property is
very important for the whole collision process, since it is in the core of the
collision that occurs a transference of energy from translational
mode to the vibrational one. If the gap increases, it turns more difficult
the energy exchange between the two modes, meaning a disfavoring of the
occurrence of bion states and n-bounce collisions with $n\ge 2$ and the
enlargement of the range of velocities with one-bounce collisions.


\section{Numerical Results}


Here we will describe our main results concerning to the number of bounces
as a function of the initial kink velocity in a symmetric kink-antikink
collision. We will consider the standard $\phi^4$ theory and twin theories
with increasing parameters $1/M^2$. Our results will be confronted with the
theoretical predictions attained previously in this paper.


\subsection{$\protect\lambda\protect\phi^4$ theory}

\begin{figure}[tbp]
\includegraphics[{angle=0,width=10cm}]{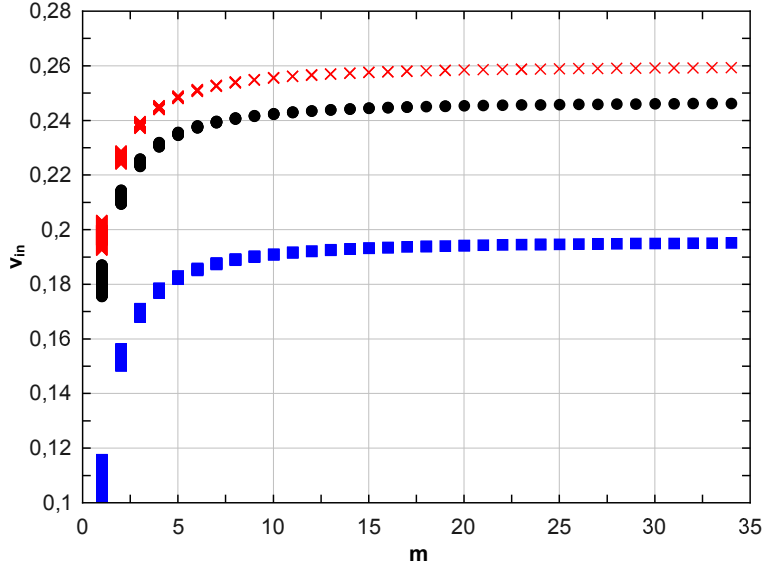}
\caption{Intervals in velocities inside which the pair kink-antikink escape after a two-bounce collision
as a function of $m$, the number of small oscillations of $\phi(0,t)$ near vacuum between the two-bounces a) for $\protect\phi^4$ model (red crosses),  b) for twin model with $1/M^2=0.01$ (black circles) and  c) for twin model with $1/M^2=0.05$ (blue squares).}
\label{phitwin_vm}
\end{figure}

First of all we review the $\lambda\phi^4$ theory. The equation of motion is 
\begin{equation}
\ddot\phi-\phi^{\prime \prime }+V_\phi=0
\end{equation}
where the dots and primes mean derivatives with respect to $t$ and $x$,
respectively. We studied a symmetric $K\bar K$ collision, with an initial
configuration where the pair $K\bar K$ is sufficiently separated for the
free solution to be useful as an initial condition (kink with velocity $%
v_{in}$, antikink with velocity $-v_{in}$). This means to chose as the
initial conditions 
\begin{eqnarray}
\phi(x,0)&=&\phi_K(x+x_0,v_{in},0)-\phi_K(x-x_0,-v_{in},0)-1 \\
\dot\phi(x,0)&=&\dot\phi_K(x+x_0,v_{in},0)-\dot\phi_K(x-x_0,-v_{in},0).
\end{eqnarray}
We used a pseudospectral method on a grid with $2048$ nodes and periodic
boundary conditions. We fixed $x_0=15$ as the initial kink position and we
set the grid boundaries at $x_{max}=120$. 
\begin{figure}[tbp]
\includegraphics[{angle=0,width=5cm}]{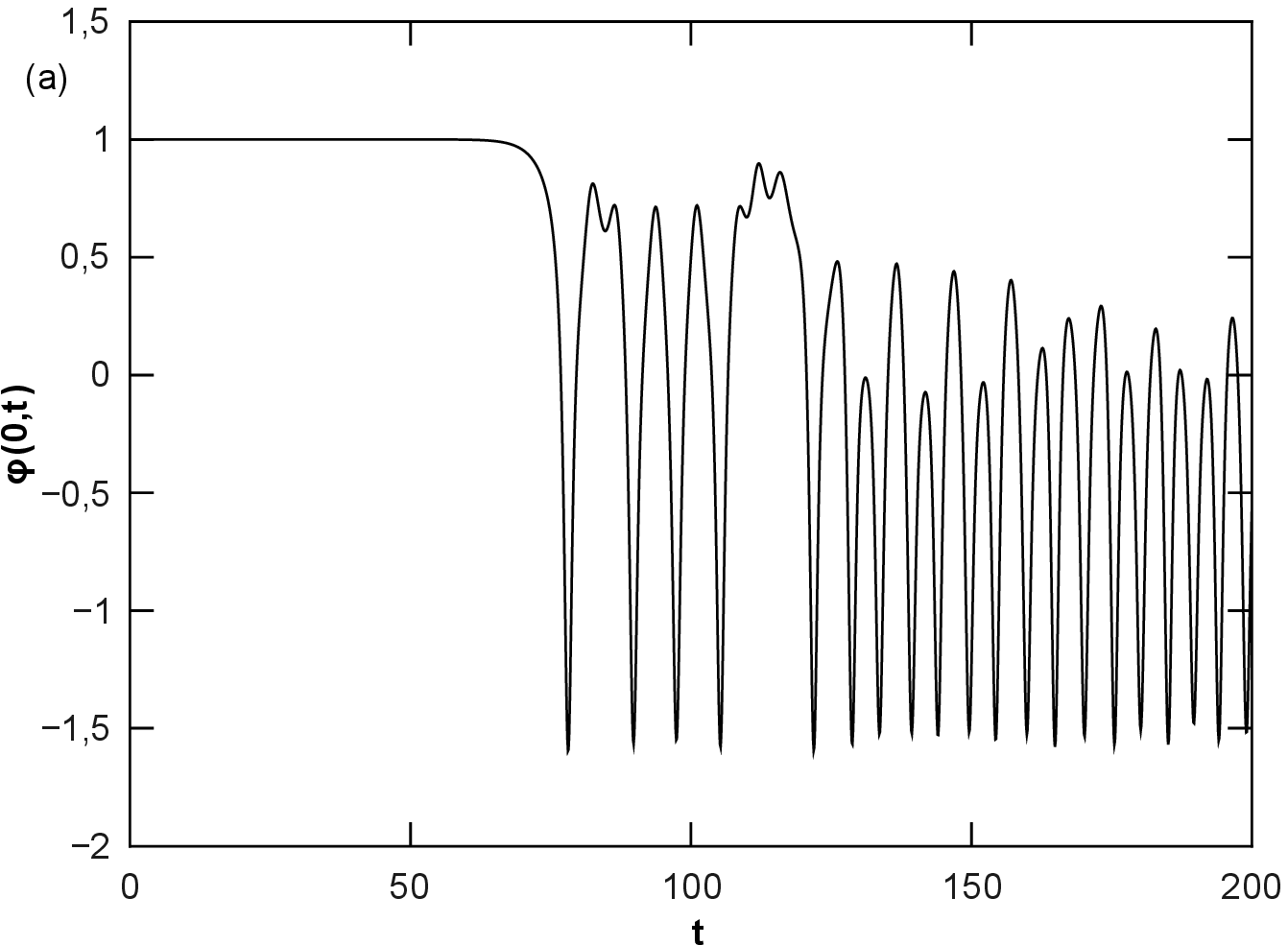} 
\includegraphics[{angle=0,width=5cm}]{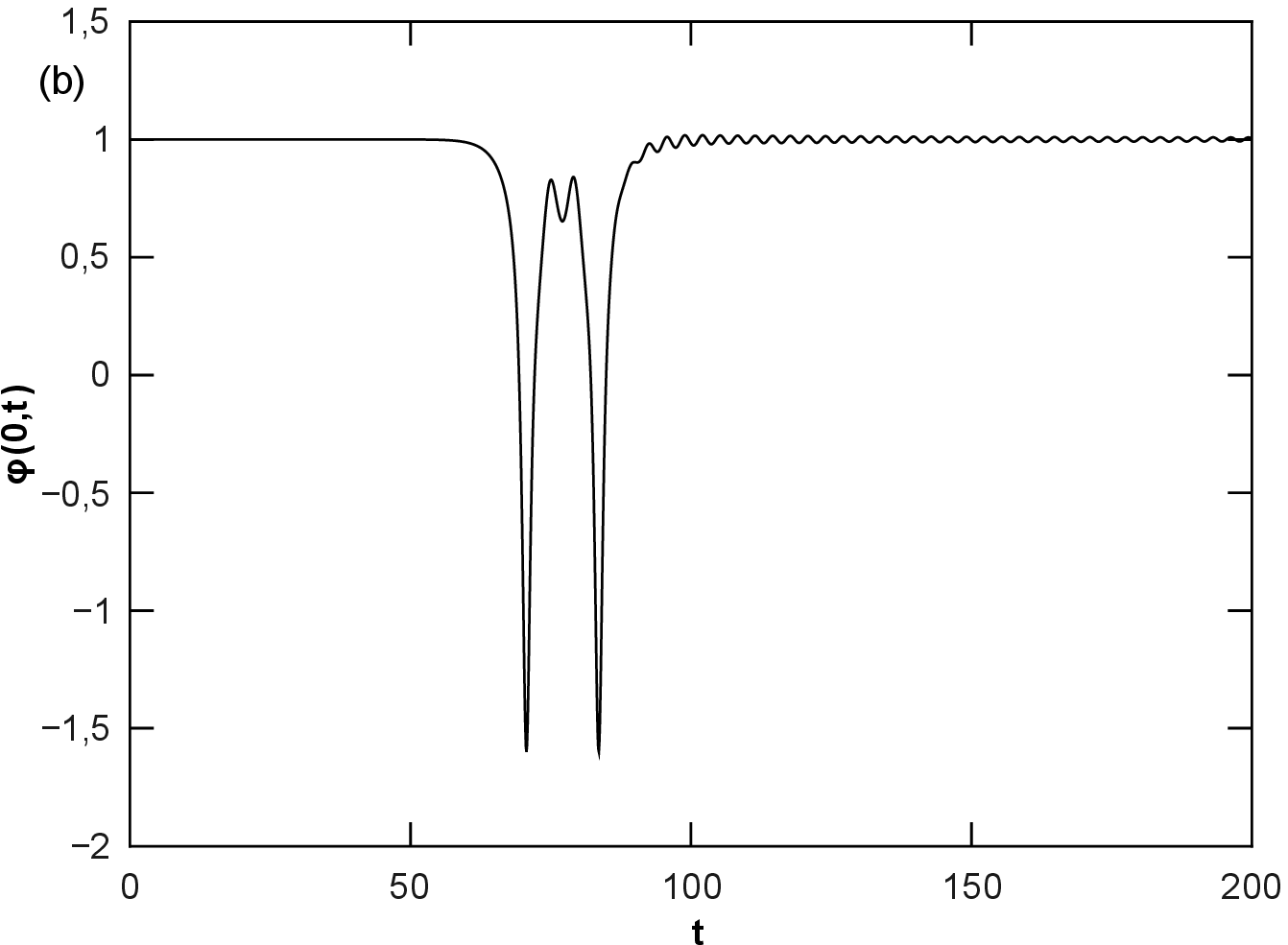}
\includegraphics[{angle=0,width=5cm}]{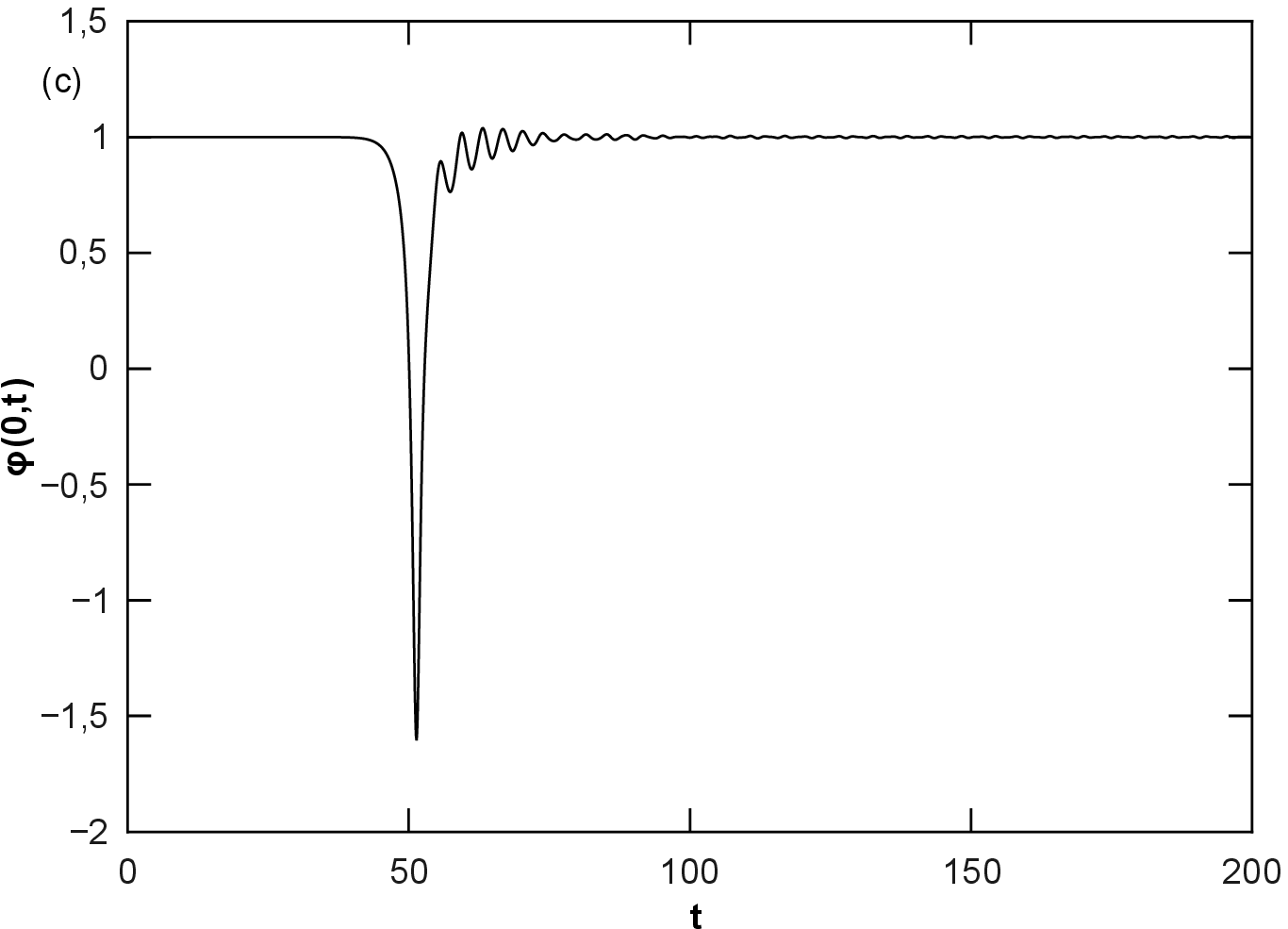}
\caption{Scalar field $\protect\phi(0,t)$ at the center-of-mass versus t for
a) $v_{in}=0.18$ (bion state), b) $v_{in}=0.20$ (two-bounce collision) and
c) $v_{in}=0.28$ (one-bounce collision). }
\label{phicm}
\end{figure}
For the $\phi^4$ model we reproduced some results from the literature \cite%
{aom} concerning to the appearance of two-bounce windows. For a particular
initial velocity $v_{in}$, the structure of bounces can be easily verified
with a plot of the scalar field at the center of mass $\phi(0,t)$ as a
function of $t$. Some examples for three different initial velocities can be
seen in Figs. \ref{phicm}, where we have a bion (\ref{phicm}a), two-bounce (%
\ref{phicm}b) and one-bounce (\ref{phicm}c) collisions. As one knows, the
dependence of the number of bounces with the modulus of the initial velocity 
$v_{in}$ of the pair kink-antikink is intricate. For low $v_{in}$ a bion
state is formed, whereas for high $v_{in}$ one has a one-bounce scattering.
For intermediate velocities, there appears two-bounce windows of variable
size separated by regions of bion states. 
\begin{figure}[tbp]
\includegraphics[{angle=0,width=5cm,height=4cm}]{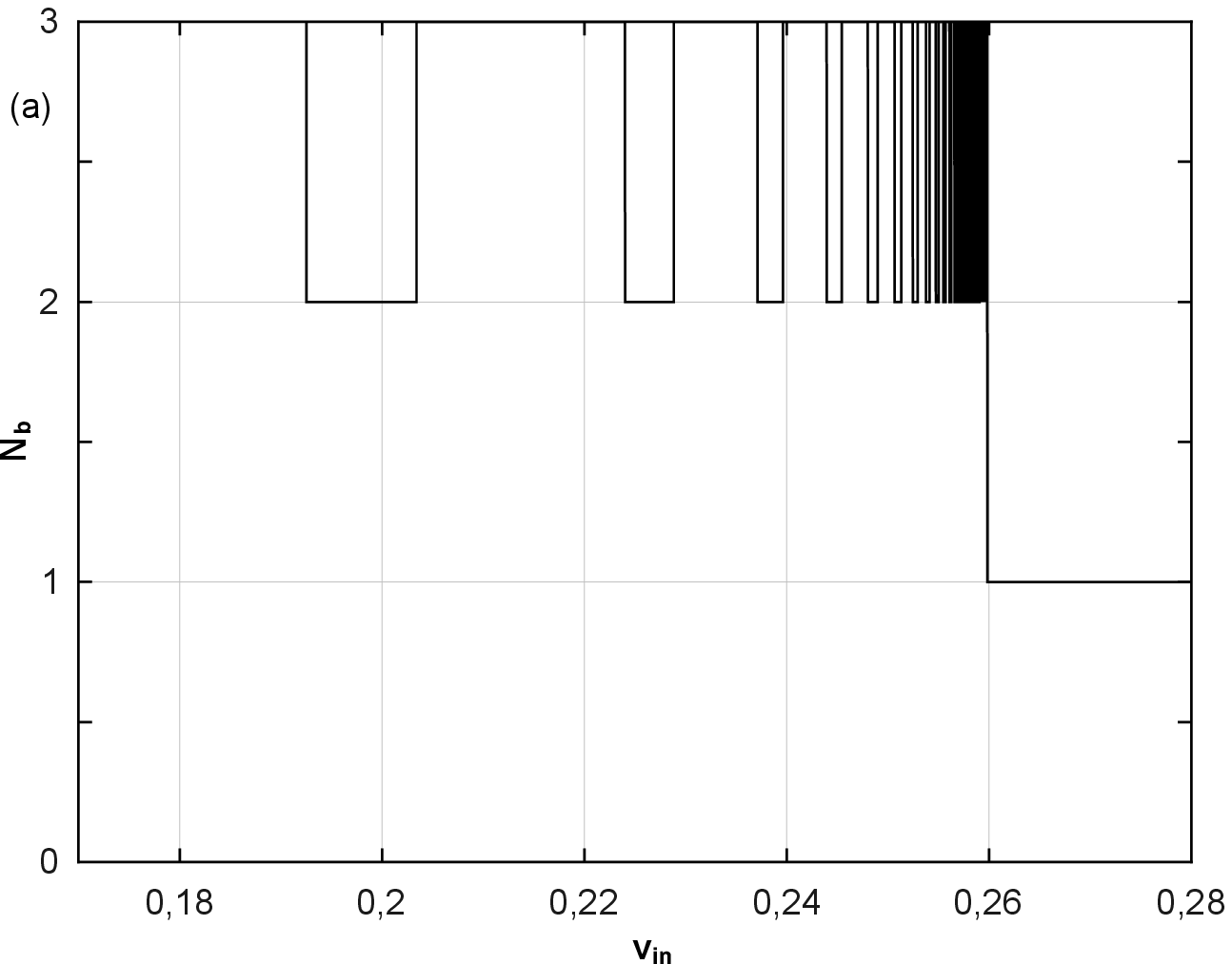}
\includegraphics[{angle=0,width=5cm,height=4cm}]{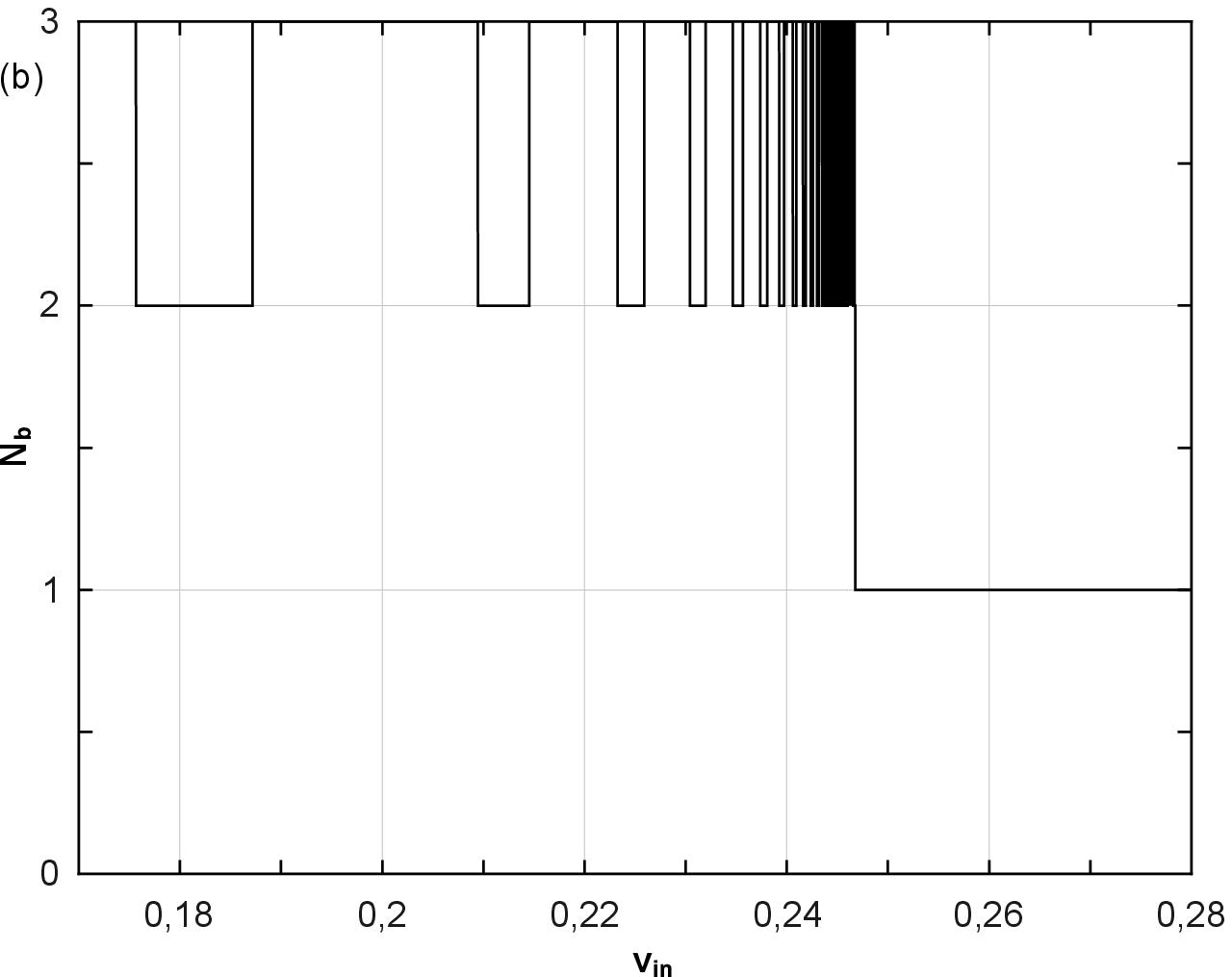}
\includegraphics[{angle=0,width=5cm,height=4cm}]{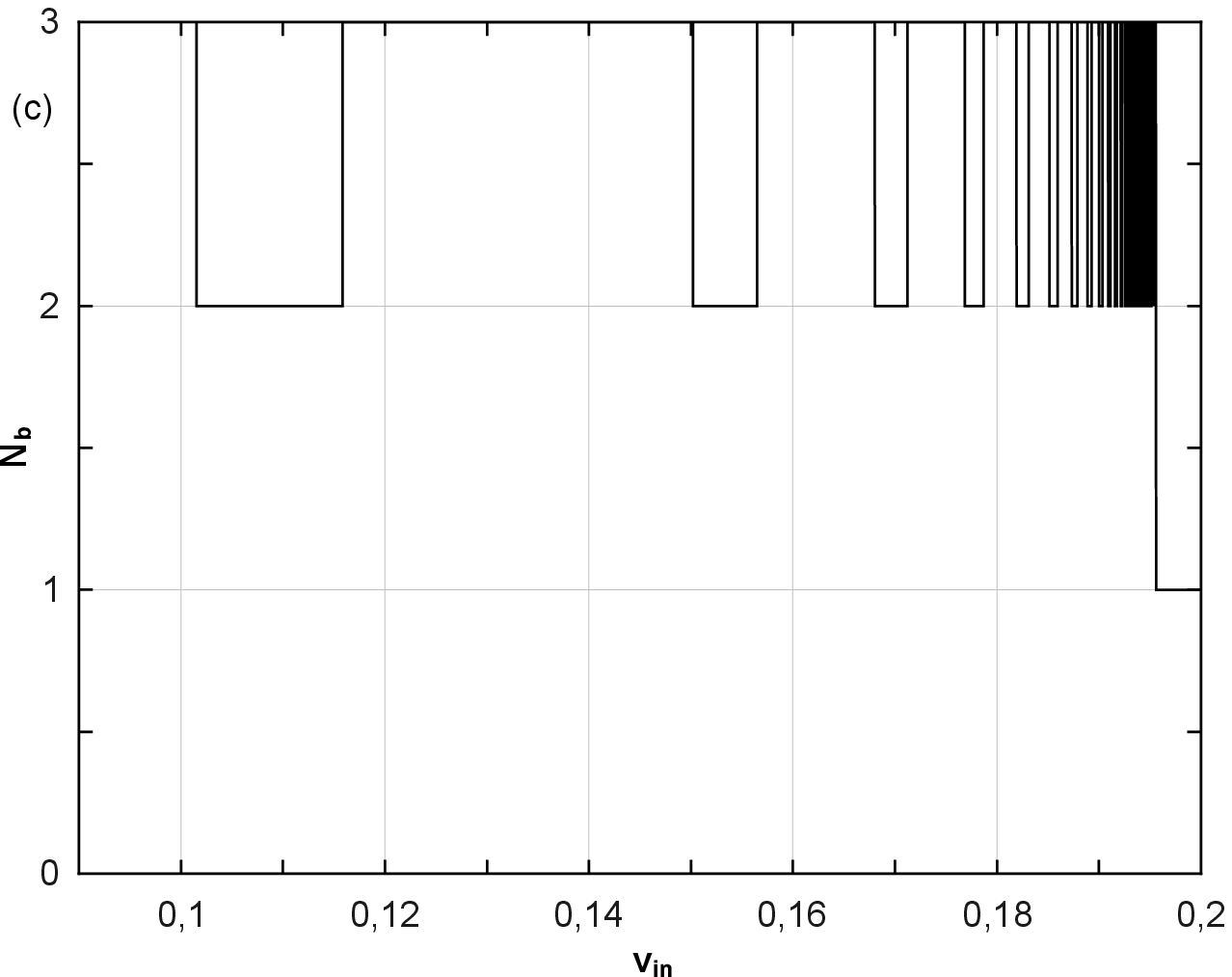}
\\
\includegraphics[{angle=0,width=5cm,height=4cm}]{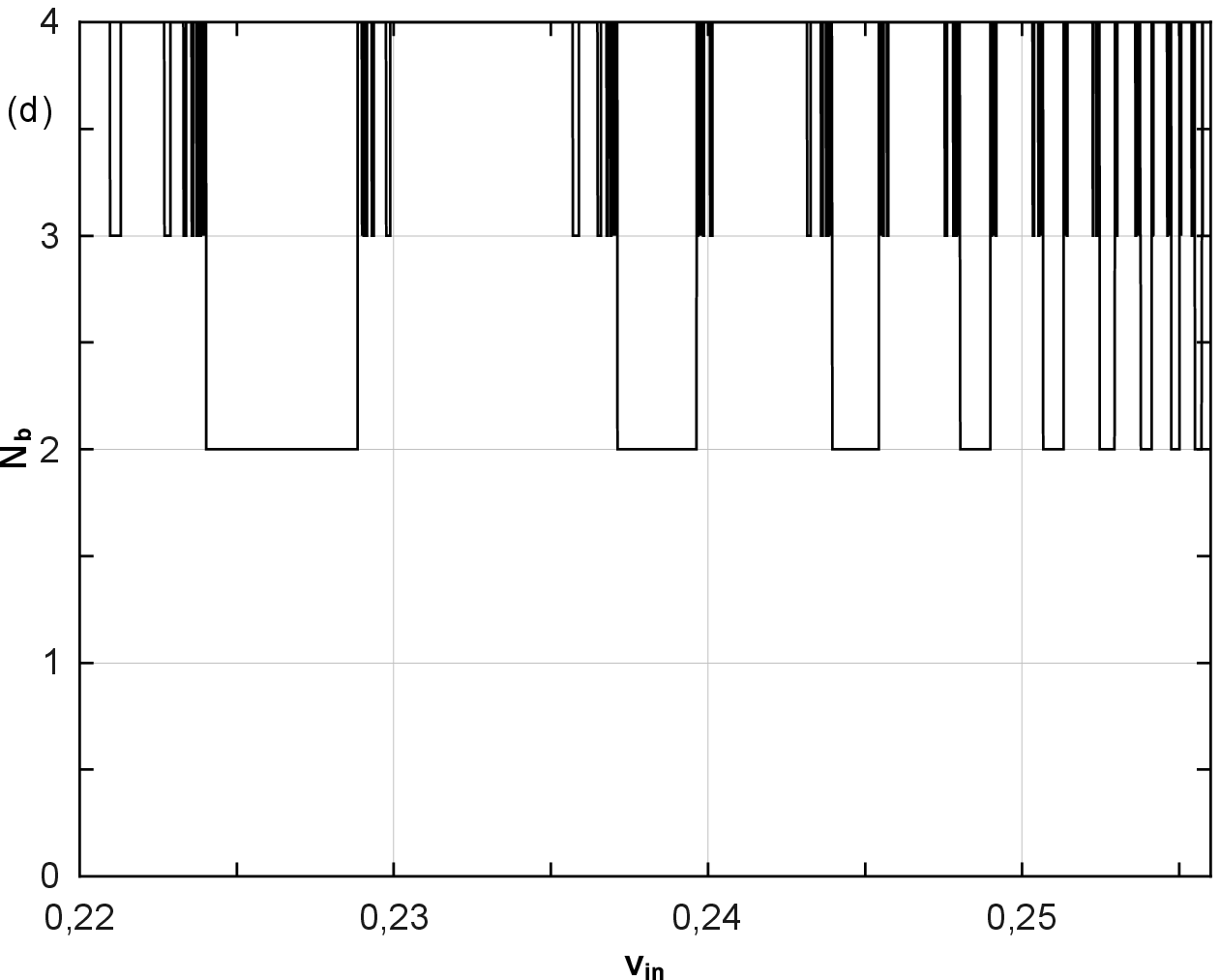}
\includegraphics[{angle=0,width=5cm,height=4cm}]{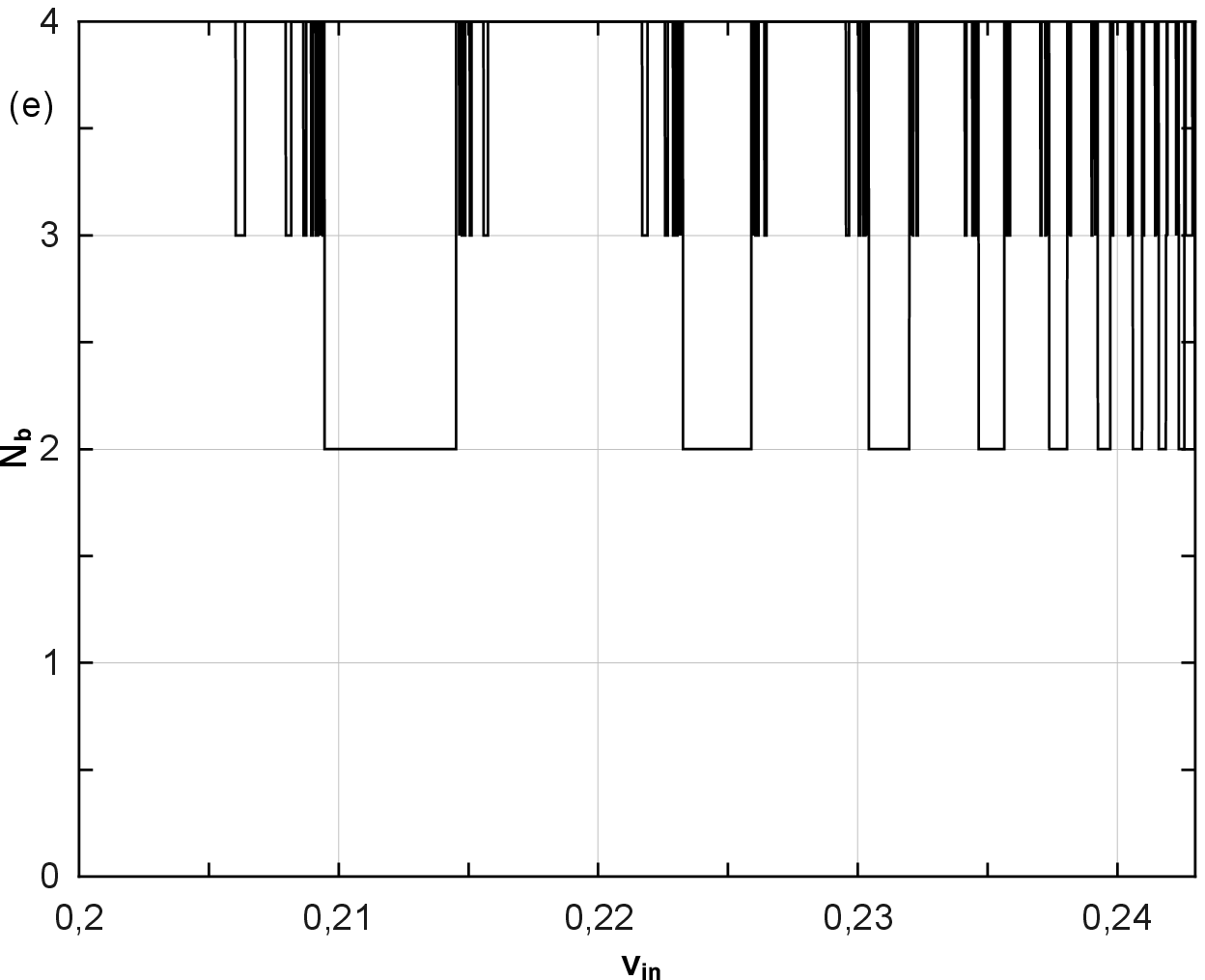}
\includegraphics[{angle=0,width=5cm,height=4cm}]{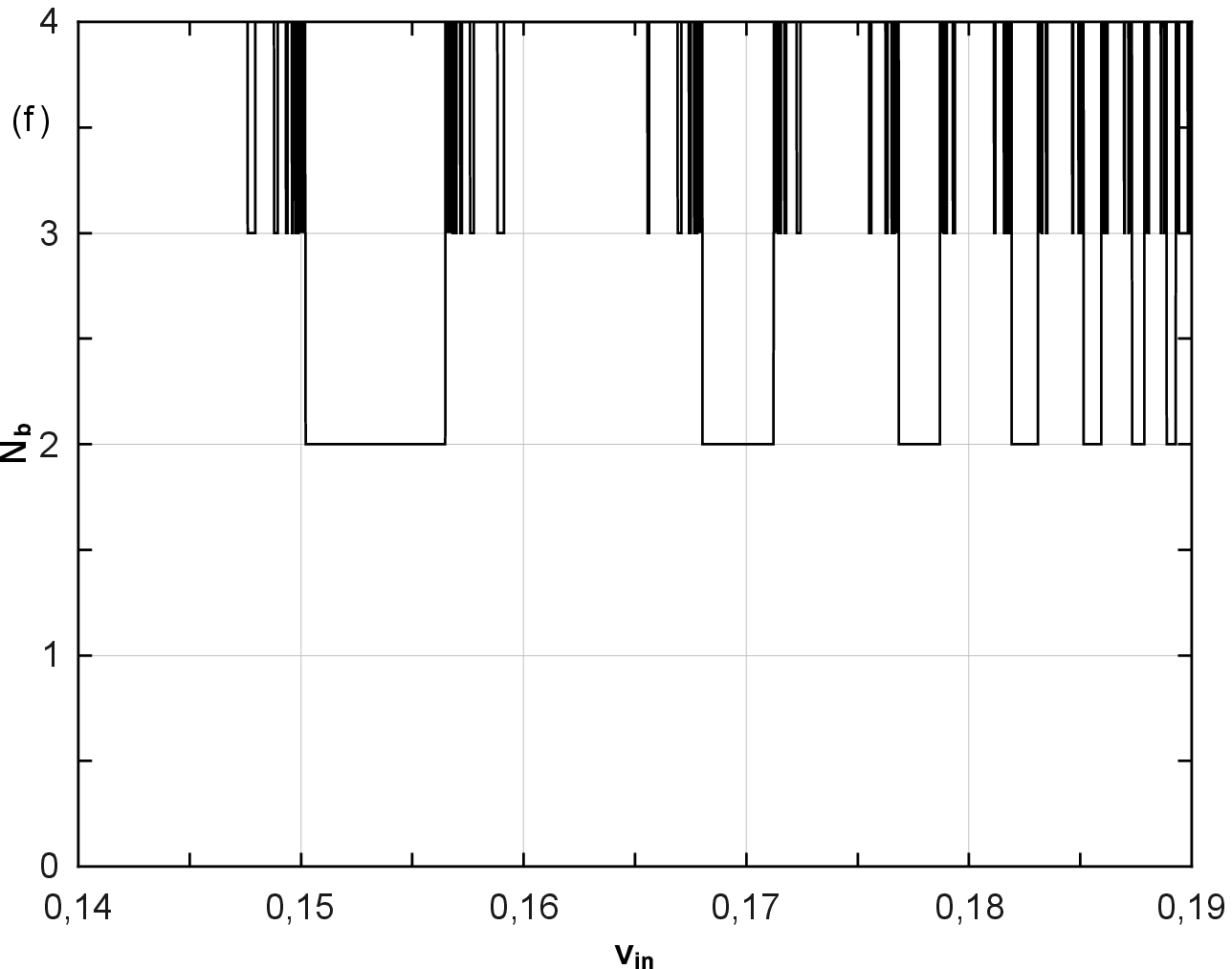}
\caption{Number of bounces versus initial velocity $v_{in}$, a) for $\protect%
\phi^4$ model (upper left) and b) for twin model with $1/M^2=0.01$ (upper center) and $1/M^2=0.05$ (upper right), showing the presence of two-bounce windows. After zooming the interval, three-bounce windows can be seen in the 
corresponding lower figures, for d) $\protect\phi^4$ model (lower left) and e) twin model with $1/M^2=0.01$ (lower center) and f) twin model with $1/M^2=0.05$ (lower right) .
}
\label{bouncv}
\end{figure}
We constructed a procedure to identify the number of bounces $N_b$ for a
given collision process in a time interval $0<t<T=200s$. We defined a bounce
as connected to the change of sign of the center-of-mass solution $\phi(0,t)$%
. In this way, Fig. \ref{phicm}b and \ref{phicm}c show collisions with $%
N_b=2 $ and $N_b=1$, respectively and \ref{phicm}a shows a bion state. In
our analysis, a too large value of $N_b$ will correspond to a bion state.
Fig. \ref{bouncv}a shows the behavior of $N_b$ as a function of $v_{in}$.
Note from the figure the change of pattern for collisions around $v_{in}\sim
0.26$. The presence of higher number of bounces for $v_{in} \lesssim 0.26$
characterizes bion states, whereas there are intermediate regions with $%
N_b=2 $. States with $N_b=1$ appear for $v_{in} \gtrsim 0.26$. The structure
of the two-bounce windows can be characterized by the integer $m$ labeling
the number of oscillations in $\phi(0,t)$ between the bounces. For example,
in the Fig. \ref{bouncv}a the first two-bounce window corresponds to $m=1$
(see details in Ref. \cite{aom}). The two-bounce thickness decreases with $m$
accumulating around $v_{in}\simeq 0.26$, the limit above which the initial
velocity is already sufficiently large for a one-bounce scattering. Table II
shows some characteristics of the thickness of the first four two-bounce
windows. This table shows how an increasing in $m$ reflects in the reducing of the
two-bounce windows. This can be better seen in Fig. \ref{phitwin_vm} (see Ref. \cite{bk}).

\begin{table}[tbp]
\begin{tabular}{|c|ccc||ccc|c|}
\hline\hline
m & $v_1$ & $v_2$ & $\Delta v$ & $v_{1T}$ & $v_{2T}$ & $\Delta {v_T}$ &  \\ 
\hline\hline
1 & 0.1926 & 0.2034 & 0.0108 & 0.1757 & 0.1872 & 0.0115 &  \\ \hline
2 & 0.2241 & 0.2288 & 0.0047 & 0.2095 & 0.2145 & 0.005 &  \\ \hline
3 & 0.2372 & 0.2396 & 0.0024 & 0.2233 & 0.2259 & 0.0026 &  \\ \hline
4 & 0.2440 & 0.2454 & 0.0014 & 0.2305 & 0.2320 & 0.0015 &  \\ \hline
\end{tabular}%
\caption{Separation in velocities of the first four two-bounce windows. $m$
is a label corresponding to the particular two-bounce window (equal to the
number of oscillations of $\protect\phi(0,t)$ between the two bounces). For
the $\protect\phi^4$ model, the columns $v_1$ and $v_2$ correspond to the
first and last points of the velocity of the window and $\Delta v$ the
corresponding width (the numerical results are in agreement with Table II of 
\protect\cite{cp}). For the twin k-defect model with $1/M^2=0.01$, columns
to the right show similar informations identified by $v_{1T}$, $v_{2T}$ and $%
\Delta v_T$.}
\end{table}

Also refining the input data
of initial velocities around a border between a two-bounce window and bion, one sees
the presence of a cascate of three-bounce windows, as can be seen in Fig. \ref{bouncv}d.
Note that the three-bounce windows accumulate around the border of the two-bounce windows, 
replicating the effect of reducing of thickness and separation of windows that occurs in Fig. \ref{bouncv}a,
and showing the well-known fractal pattern of $K\bar K$ collisions for the $\phi^4$ model.


\subsection{twin theory}


The twin theory in the regime $1/M^2\ll1$ has the corresponding equation of
motion 
\begin{equation}  \label{eom_twin}
\ddot\phi-\phi^{\prime \prime }+\frac1{M^2}(\dot\phi^2\ddot\phi+\phi^{\prime
2}\phi^{\prime \prime }-2\dot\phi\phi^{\prime }\dot\phi^{\prime
})+U_\phi-\frac1{M^2}UU_\phi=0.
\end{equation}
In the regime $1/M^2\ll1$, and neglecting terms of ${\mathcal{O}}(1/M^4)$ we
have 
\begin{equation}
U(\phi)=\frac12W_\phi^2-\frac18 \frac{W_\phi^4}{M^2}.
\end{equation}
\begin{figure}[tbp]
\includegraphics[{angle=0,width=5cm}]{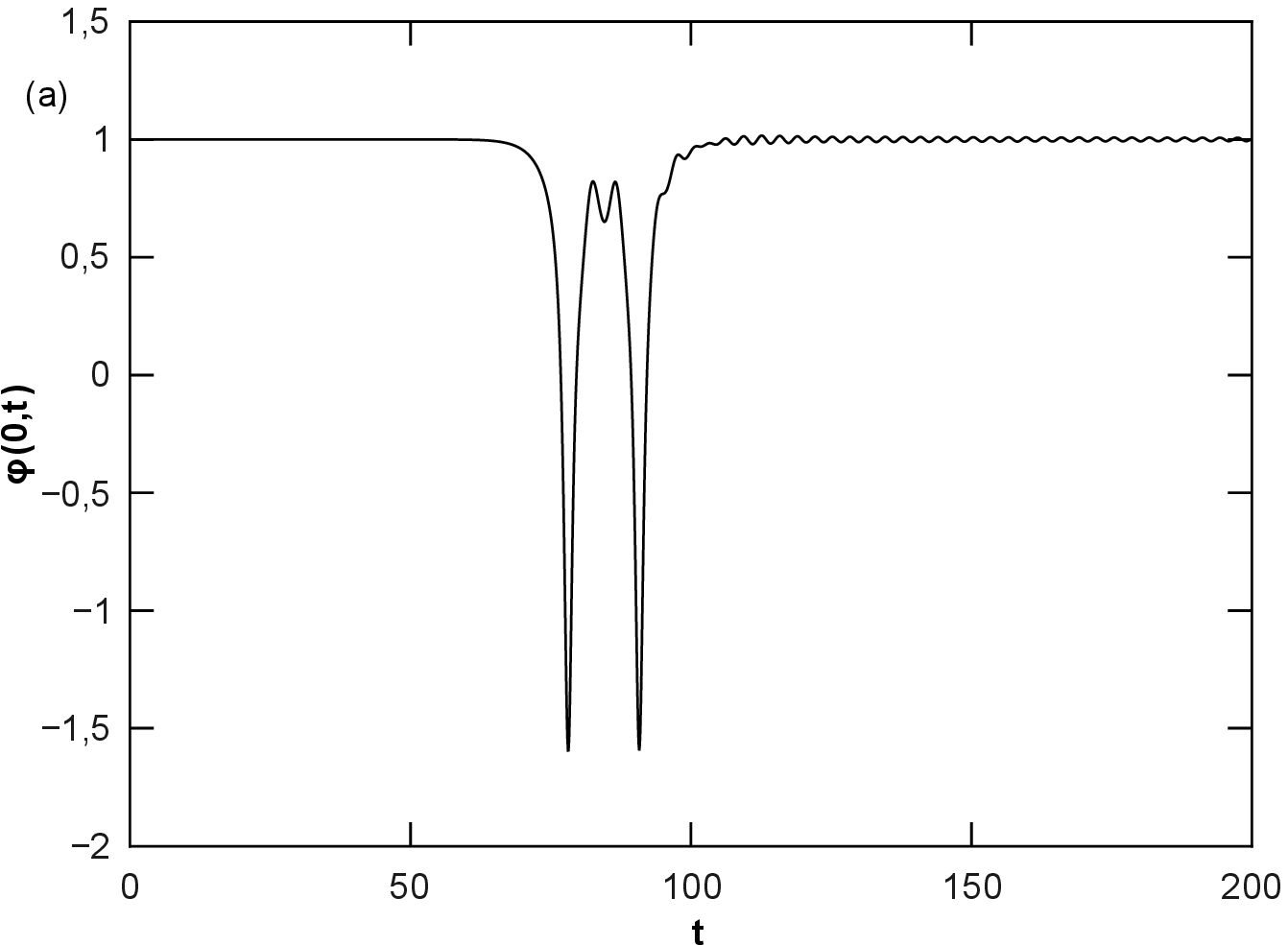}
\includegraphics[{angle=0,width=5cm}]{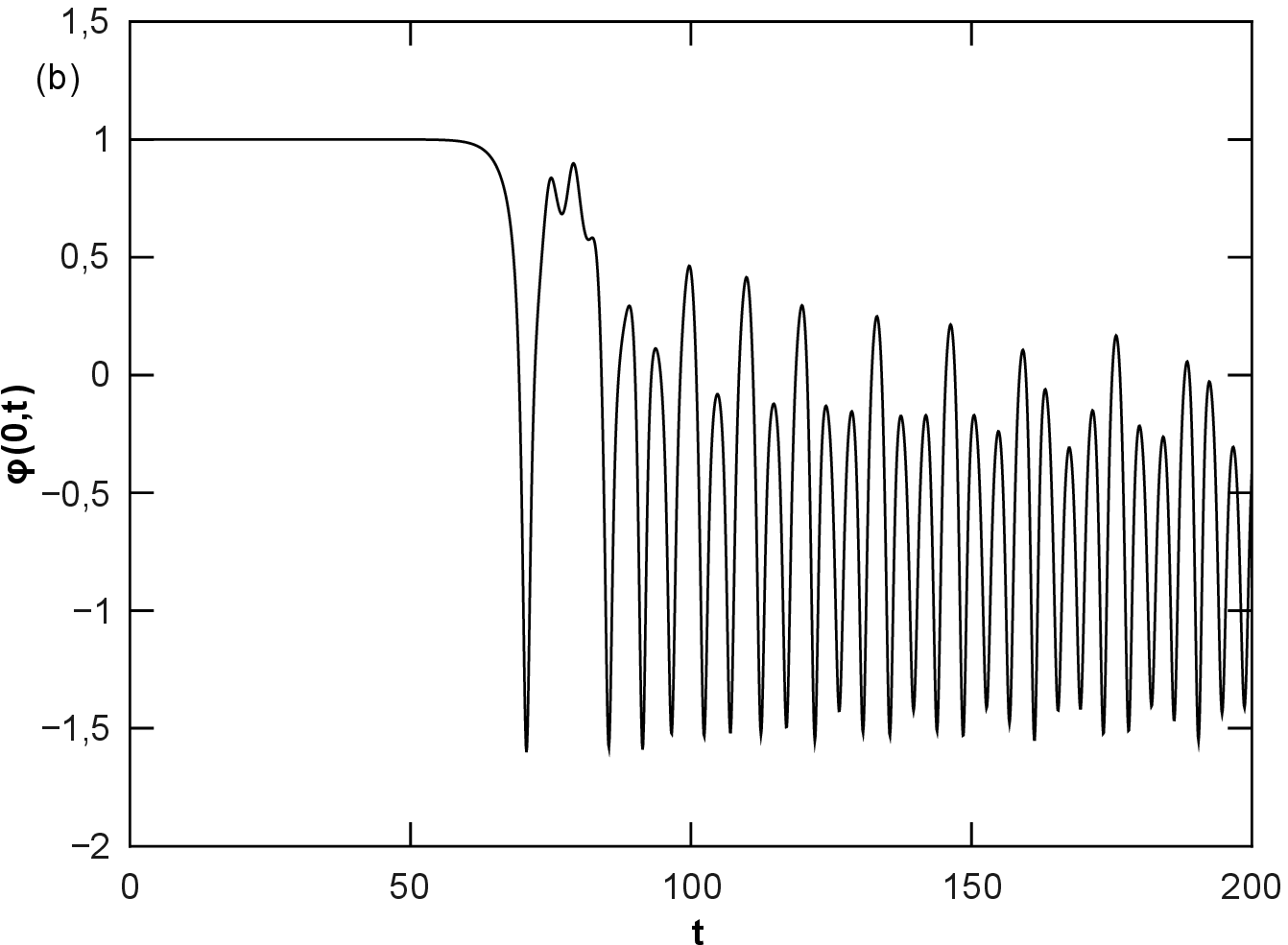}
\includegraphics[{angle=0,width=5cm}]{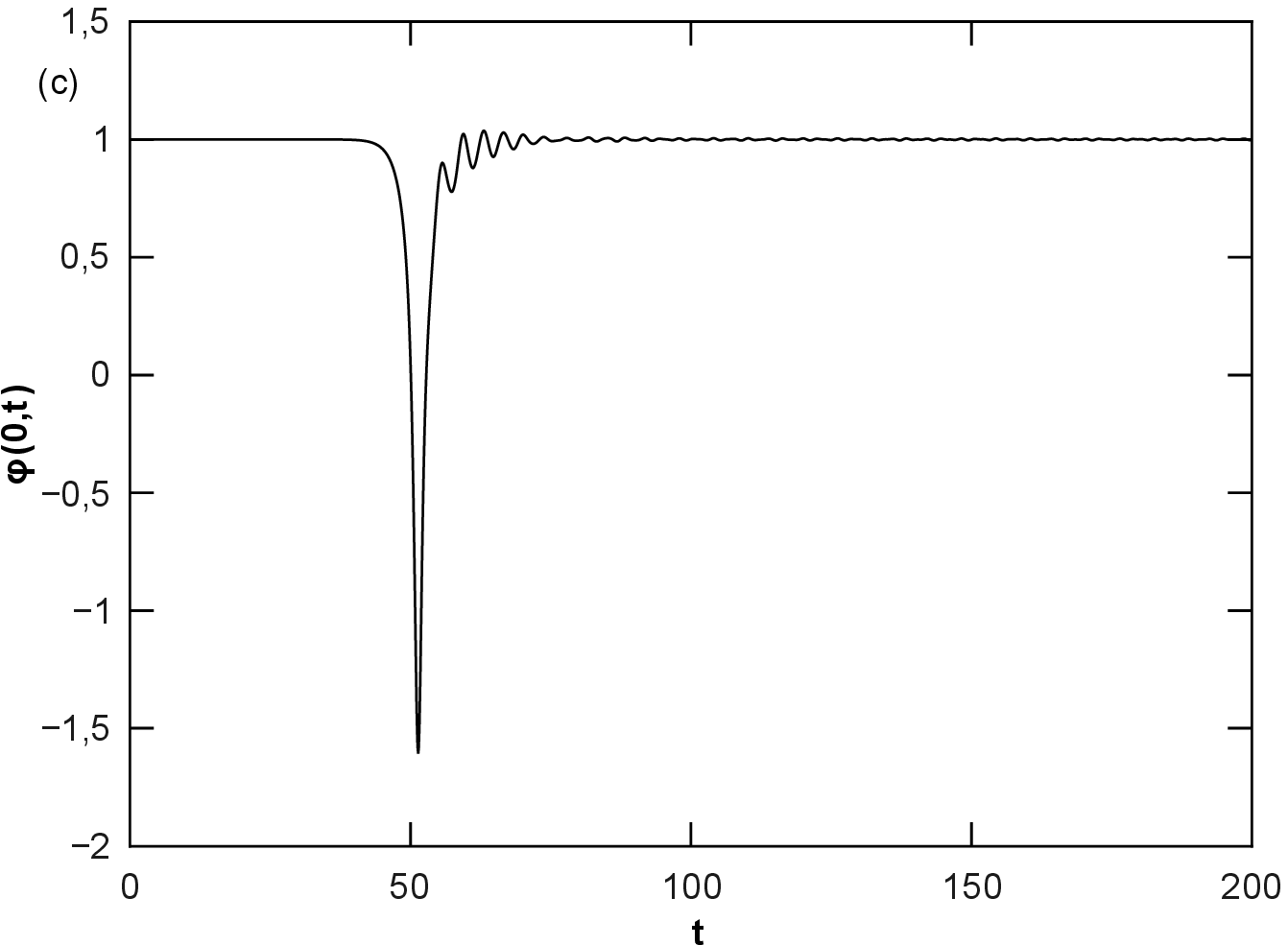}
\caption{Scalar field $\protect\phi(0,t)$ at the center-of-mass versus t for
a) $v_{in}=0.18$ (two-bounce collision), b)$v_{in}=0.20$ (bion state) and c) 
$v_{in}=0.28$ (one-bounce collision) for twin model with $1/M^2=0.01$. }
\label{phitwin}
\end{figure}
Substituting this in Eq. (\ref{eom_twin}), one can easily verify that $%
\phi_K(x,t)=\tanh(\gamma(x-vt))$ is the solution for a free propagating
kink. As expected this is the same solution already achieved for the $\phi^4$
model. This means that the initial conditions will be the same used for the $%
\phi^4$ model. Indeed, it is only the collision process that will
distinguish both theories. Also, as used previously for the $\phi^4$ theory,
this analytic solution can be used as a first test for the numerical
solution for the twin theories. In Figs. \ref{phitwin} we present some
results for $\phi(0,t)$ for the twin theory with $1/M^2=0.01$. Comparing the
figures with Fig. \ref{phicm} for the $\phi^4$ theory we see that even for a
small value of $1/M^2$ the behavior is altered sensibly. For example, for $%
v_{in}=0.2$ we have two-bounce collision with $m=1$ for the $\phi^4$ model
(Fig. \ref{phicm}b) and bion for the twin model (Fig. \ref{phitwin}b). Also,
for $v_{in}=0.18$ we have bion for the $\phi^4$ model (Fig. \ref{phicm}a)
and two-bounce collision with $m=1$ for the twin model (Fig. \ref{phitwin}%
a). Our results of $N_b$ as a function of $v_{in}$ for $1/M^2=0.01$ are
depicted in Fig. \ref{bouncv}b. The same effect of appearance of two-bounce
windows already known for the $\phi^4$ model is present for the twin model.
In the twin model, however, the one-bounce collision occurs for $%
v_{in}\gtrsim 0.246$, smaller that $v_{in}\gtrsim 0.26$ for the $\phi^4$
theory. An increasing of $1/M^2$ shows the same pattern, as can be seen in  Fig. \ref{bouncv}c for 
$1/M^2=0.05$, where now the one-bounce collisions occur for $v_{in}\simeq 0.20$. This enlargement of the region where occurs the one-bounce window is
directly related to the increasing of the gap between the translational and
vibrational modes, as discussed in the final of the previous section. 

Table II shows the separation in velocities of the first four two-bounce windows
for both $\phi^4$ and the twin model for $1/M^2=0.01$. We see that the thickness of the windows
are roughly the same, occurring for lower velocities in the twin model in
comparison to the $\phi^4$ model. This can be better seen in Fig. \ref{phitwin_vm}, where we also included the case where $1/M^2=0.05$. There one can also see that, similarly to the $\phi^4$ model, for the twin model the velocity thickness is reduced continuously with $m$, whereas $v_{in}$ grows and asymptotes to the minimum velocity $v_{in}^*$ for the occurrence of a one-bounce collision. Also note from the figure that the larger is $1/M^2$, the lower is $v_{in}^*$.

For the twin model we refined the input data
of initial velocities around a border between a two-bounce window and bion, 
obtaining the results of Figs. \ref{bouncv}e and \ref{bouncv}f. There one can see the presence of a cascate of three-bounce windows, accumulating around the border of the two-bounce windows, 
replicates the effect of reducing of thickness and separation of windows that occurs in the corresponding Figs. \ref{bouncv}b and \ref{bouncv}c,
and showing that there is a fractal pattern of $K\bar K$ collisions for the twin model in a similar way to the verified for $\phi^4$ model (compare with Figs. \ref{bouncv}a and \ref{bouncv}d).  

\begin{figure}[tbp]
\includegraphics[{angle=0,width=8cm,height=6cm}]{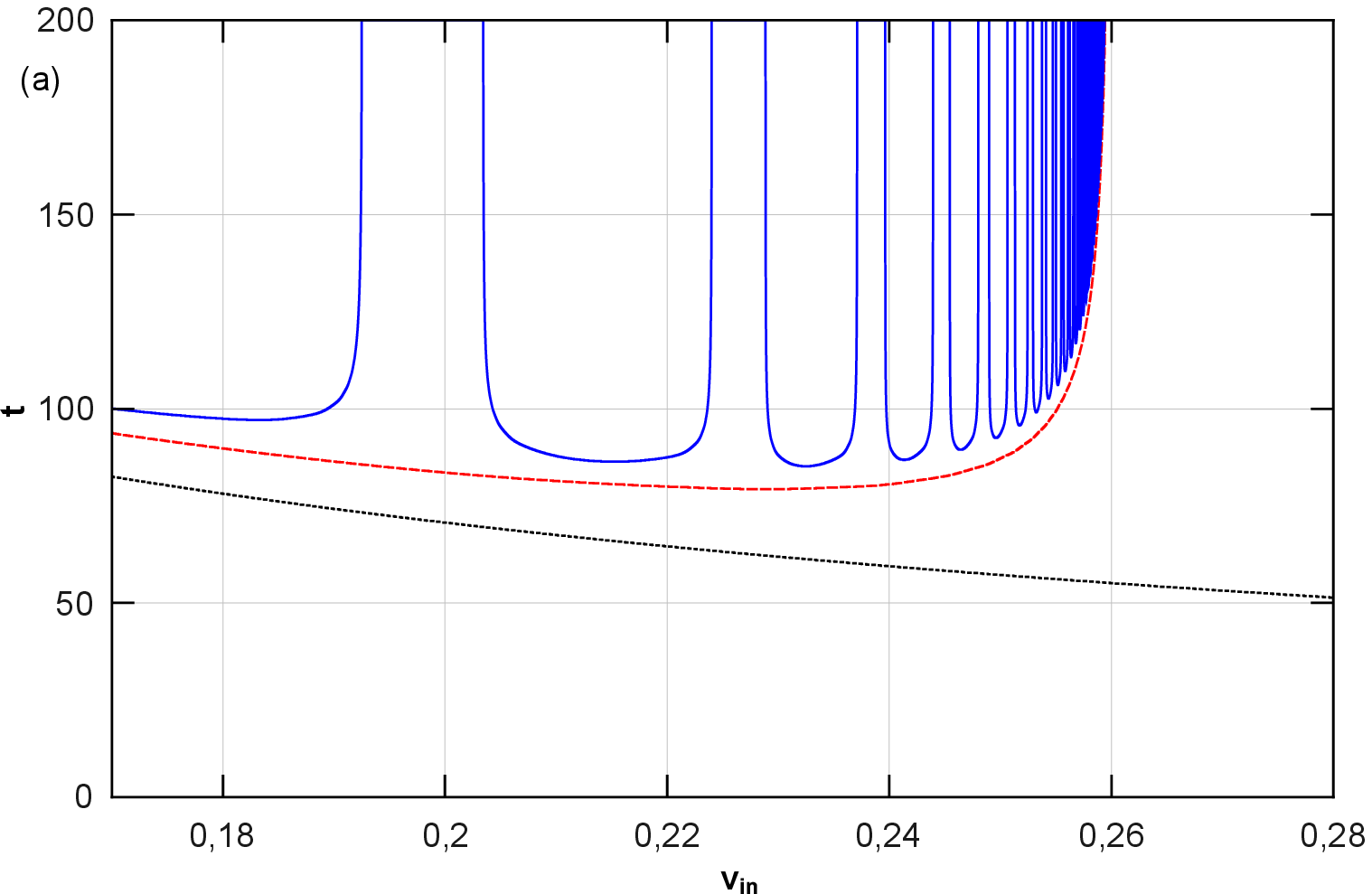}
\includegraphics[{angle=0,width=8cm,height=6cm}]{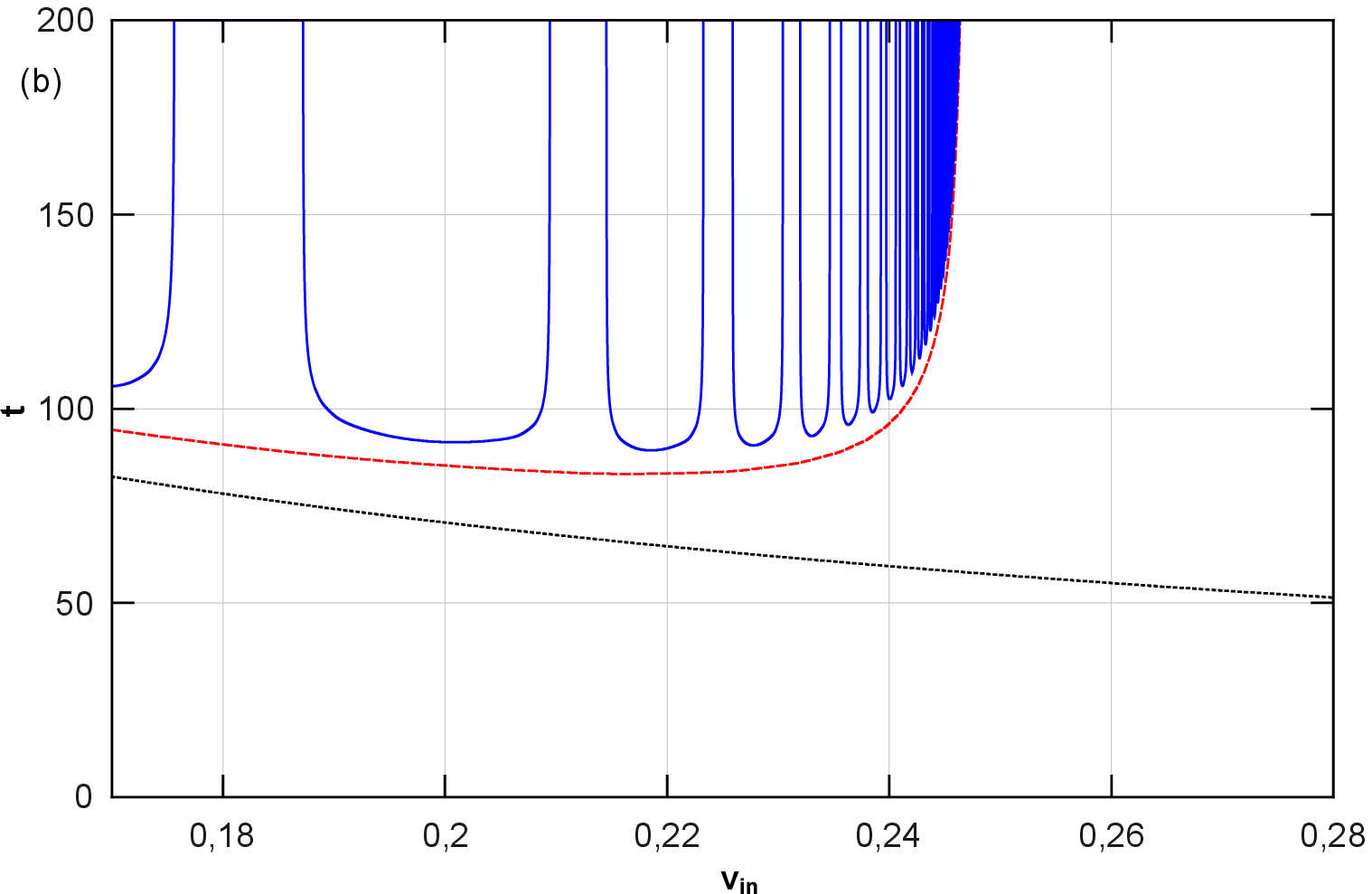}
\caption{Times to the first (black line), second (red long trace)
and third (blue short trace) bounces in a $K\bar{K}$ collision, as a function of the initial velocity, for a) $\protect\phi^4$
model and b) twin model with $1/M^2=0.01$.}
\label{bouncv_twin}
\end{figure}

Fig. \ref{bouncv_twin}a-b show the time $t$
of the first three bounces as a function of the initial velocity $v_{in}$.
Our results with $1/M^2=0.01$ (Fig. \ref{bouncv_twin}b) clearly show that
there is a displacement of the two-bounce windows for regions of lower
velocities in comparison to the $\phi^4$ model (compare with Fig. \ref%
{bouncv_twin}a). This can be interpreted as a signal of a weaker $K\bar K$
interaction for the twin model, in comparison to the $\phi^4$ model.

\begin{figure}[tbp]
\caption{Time between bounces versus $n=m+2$, where $m$ is the window number for a) $1/M^2=0$ (dotted black), b) $1/M^2=0.01$ (traced red), c) $1/M^2=0.05$ (dot-traced blue) and d) $1/M^2=0.08$ (solid green). For fixed $1/M^2$ circles represents numerical results, whereas line is a least-square fitting.}
\includegraphics[{angle=0,width=8cm,height=6cm}]{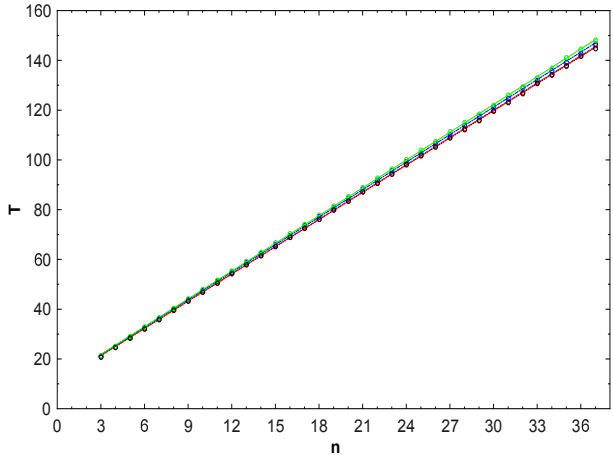}
\label{slope_twin}
\end{figure}

We also investigated if the fundamental relation given by Eq. (\ref{T_omega}) holds in the modified twin model.
In Fig. \ref{slope_twin} we plot the time between bounces versus $n=m+2$, where $m$ is the window number. The integer $n$ is constructed in order to achieve a phase shift $\delta$ between $0$ and $2\pi$. We note from the figure that the numerical points can be described with good approximation by a straight line. Table III compares the angular coefficients from the least-squares with the theoretical one $2\pi/\omega_1$, predicted by Eq. (\ref{T_omega}). From the Table we see that for $1/M^2=0$ the theoretical angular coefficient $2\pi/(\omega^{(1)}_1)$ is lower but comparable to the numerical value obtained by least-squares method, as already noted in Ref. \cite{aom}. Also Table III shows that the numerical angular coefficient {\it grows} with the increasing of $1/M^2$. This is an intriguing character, since from Eq. (\ref{T_omega}) it is expected a {\it decreasing} of the angular coefficient, as shown in the second row of Table III. 
Since we found no physical ground for such behavior, we must say this numerical result must be handled with care, for the following reasons: i) One must note that Fig. \ref{slope_twin} considers a quite large number of windows (up to $m\sim 35$). As the window number grows, also grows the numerical error of the simulations. To put in other words, a consistent numerical analysis of two-bounce windows is easier from the computational perspective. 
ii) Eq. (\ref{T_omega}) was proposed by Campbell et all in Ref. as a good simplification of an intrincate process of interaction between two extended objects. If we want to test a qualitative expression, we must not be so strict in quantitative agreement.
iii) We cannot consider a too large value of $1/M^2$ as a valid approximation of a twin theory breaks. We could take $1/M^2=0.05$ as a higher bound for an analysis of a two-bounce windows, but this value could be even lower for larger number of bounces.
In this paper, all the former numerical results and analysis concerning to two-bounce windows where shown to be compatible with the interpretation of Eq. (\ref{T_omega}). The aplicability of this equation can be confirmed also by Fig. \ref{slope_twin} and Table III in the sense that a straight line behavior was verified and the angular coefficients are roughly comparable, as already done in Ref. \cite{aom} for the $\phi^4$ model.

\begin{table}[tbp]
\begin{tabular}{|c|c|c|c|}
\hline\hline
$1/M^2$ & $2\pi/(\omega^{(1)}_1)$ & $a$ & relative error \\ 
\hline\hline
0 & 3.63 & 3.65 & $0.55 \% $\\ \hline
0.01 & 3.62 &  3.66 & $1.1 \% $\\ \hline
0.05 & 3.58 & 3.69 & $3.0 \% $ \\ \hline
\end{tabular}%
\caption{Theoretical angular coefficient $2\pi/(\omega^{(1)}_1)$ versus angular coefficient $a$ obtained by least-squares method from Fig. \ref{slope_twin}.}
\end{table}

\section{conclusions}

In this work we have studied kink-antikink collisions for twin theories. We
were particularly interested in investigating in which aspect the presence of
a general kinetic content (k-generalization) in the Lagrangian could be
revealed in a collision process. Starting form a general Lagrangian $%
\mathcal{L}(X,\phi)$, and considering a convenient decomposition of the
fluctuations, we analyzed the energy contribution of the fluctuations. 
After reviewing the first-order
formalism of twin theories, we considered the $\phi^4$ theory and a class of
twin theories depending on a mass parameter $M$. In the regime where $%
1/M^2\ll 1$, we obtained the spectra of excitations with two bound states: a
zero-mode, responsible for the translation, and a vibrational one, crucial
for the two-bounce collisions. We showed that the gap between the two bound
states is larger for the twin model, and that it increases with $1/M^2$. A
detailed numerical analysis reproduced some known results for $K\bar K$
collisions in the $\phi^4$ theory, used as a control model to be confronted
with the results of a twin model of a general kinetic content. The numerical
results corroborated the theoretical expectation that, in a collision
process, the increasing of $1/M^2$ reduces the possibility of formation of a
trapped $K\bar K$ bion state.

\section{Acknowledgements}

The authors thank FAPEMA, CAPES, CNPq and IFMA for financial support. The
authors thank Herbert Weigel and R. Casana for clarifying several points of stability
analysis. A. R. Gomes thanks A. S. Anjos and M. M. Ferreira Jr.
for discussions.

\end{document}